\documentclass[12pt]{iopart}

\usepackage{graphicx}
\usepackage{latexsym}
\usepackage{amsfonts,ulem,cancel}
\usepackage{graphicx}
\usepackage{fullpage}
\usepackage{esdiff}
\usepackage{tocloft}

\usepackage{fancybox}
\usepackage{color}
\usepackage{paralist}
\usepackage{flafter}
\usepackage{placeins}
\usepackage{tabularx}

\begin{document}

\title[Air fluidized balls in a background of smaller beads]{Air fluidized balls in a background of smaller beads}

\author{M. E. Beverland, L. J. Daniels and D. J. Durian}
\address{Department of Physics \& Astronomy, University of Pennsylvania, Philadelphia, Pennsylvania 19104-6396, USA}

\begin{abstract}
We report on quasi-two-dimensional granular systems in which either one or two large balls is fluidized by an upflow of air in the presence of a background of several hundred smaller beads. A single large ball is observed to propel ballistically in nearly circular orbits, in direct contrast to the Brownian behavior of a large ball fluidized in the absence of this background. Further, the large ball motion satisfies a Langevin equation with an additional speed-dependent force acting in the direction of motion. This results in a non-zero average speed of the large ball that is an order of magnitude faster than the root mean square speed of the background balls.  Two large balls fluidized in the absence of the small-bead background experience a repulsive force depending only on the separation of the two balls.  With the background beads present, by contrast, the ball-ball interaction becomes velocity-dependent and attractive. The attraction is long-ranged and inconsistent with a depletion model; instead, it is mediated by local fluctuations in the density of the background beads which depends on the large balls' motion.
\end{abstract}


\section{Introduction}

A major challenge of modern physics is to understand the behaviour of non-thermal systems \cite{Reulle2004, CMMP2010}, such as granular media \cite{Nedderman, BehringerRev, Duran}.  Just as for their thermal counterparts, these systems can display well-defined statistical distributions that encapsulate their behavior.  For thermal systems, the standard theory of statistical mechanics can be used to predict all such distributions from equations of motion based on interaction forces plus stochastic forces set by temperature \cite{Kubo1991}. For non-thermal systems, by contrast, there is no such general approach.  But in some special cases, experiments have found thermal-like behavior in the shape of distributions and in the agreement of distinctly defined effective temperatures \cite{Clark1990, Bideau1995, UrbachPRE99, PrentissAJP00, Baxter2003, Danna2003, Ojha2004, SongMakse05, Abate2005, Abate2008, MaksePRE08}.  Unfortunately there is no general framework for determining when such a thermal analogy ought to hold.  To make progress, experimentally, it is sensible to compile further examples of different kinds of driven systems where microscopic statistical distributions may be measured.

Here we report on measurements of a system consisting of a monolayer of hundreds of small beads, together with one or two large balls, fluidized in a steady upflow of air.  These beads and balls all roll on a horizontal sieve without slipping and experience in-plane forces from collisions with one another and from the air that flows up through them at high Reynolds number. Previous experiments with this apparatus have shown that a single ball \cite{Ojha2004}, as well as dense collections of many beads \cite{Abate2008}, all behave thermally.  However, for two spheres the thermal analogy is progressively broken by increasing the size disparity \cite{Abate2005}.  Here, for one large ball in a background of small beads, it is therefore unclear in advance whether or not to expect thermal behavior.  Furthermore, for two large balls in a background of small beads, it is unclear in advance whether to expect the interaction between two large balls to be repulsive due to turbulence, as in Ref.~\cite{Abate2005}, or to be attractive due to a depletion-like entropic force mediated by the thermal background of small beads.   The latter possibility of a non-equilibrium depletion force was found in Ref.~\cite{UrbachPRE07} for  a system of large balls in a background of small beads, all subject to vertical vibration.  Our motivation is both to explore these specific issues, as well as to provide a well documented experimental system to help motivate and test future statistical theories for non-thermal systems.

Our approach is based on high speed digital video, to track the sphere positions versus time.  We begin with the behavior of the one-ball system in Section~\ref{oneball}. The large ball is observed to propel ballistically through the background medium, creating compressed and rarefied regions in the background beads. We quantify these local fluctuations in the background density in Section~\ref{background}. We then characterize the large ball behavior by measuring statistical distributions, calculating the ball dynamics, and identifying its equation of motion. We find that the background generates a novel speed-dependent force on the large ball which accelerates it forward, causing it to move much faster than the small beads. In Section~\ref{twoball}, we analyze the two-ball system and calculate the same time-independent statistical distributions and dynamics for the two large balls. Lastly, in Section~\ref{interact}, we deduce the interaction between the two simultaneously-fluidized large balls, which we show to be long-ranged and attractive.  The magnitude and the range are both larger than for a depletion force.


\section{Experimental Details}\label{exp}
\begin{figure}[h!]
\begin{centering}
\includegraphics[width=0.9\textwidth]{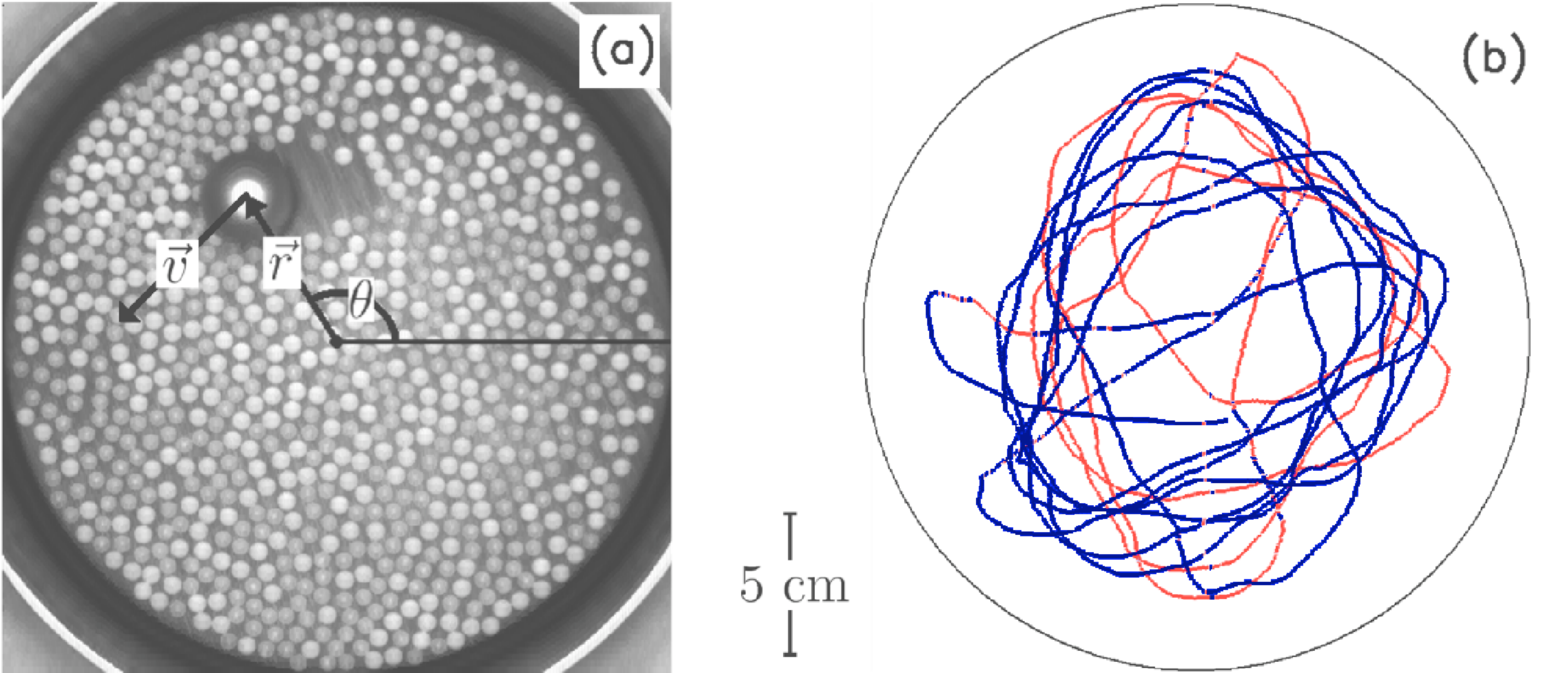}
\par\end{centering}
\caption{\label{PhotoApparatus} a) A large sphere, radius $a=1.9$~cm and mass 2.7~g, fluidized in the presence of a bidisperse background of smaller spheres -- radii 0.477~cm and 0.397~cm and masses 0.21~g and 0.165~g, respectively. The diameter ratio of the large ball to the background beads is 4.4. The radius of the system is $R=14.3$~cm. The vectors define position $\vec{r}$, velocity $\vec{v}$, as well as the polar angle $\theta$ as measured from the horizontal axis. b) A one-minute long trace of the position of the large ball shown in a). To emphasize the observed circular motion, the trace is colored red for clockwise motion, blue for counterclockwise.}
\end{figure}

The principal system that we investigate, shown in Fig.~\ref{PhotoApparatus}(a), involves a monolayer of bidisperse plastic spheres -- having radii 0.477~cm and 0.397~cm and masses 0.21~g and 0.165~g, respectively -- fluidized by an upflow of air. This monolayer constitutes the background and occupies an area fraction of 55$\%$ in the absence of larger fluidized balls. This particular area fraction was chosen so that the background is uniformly distributed across the entire system. The larger balls fluidized in the presence of this background are ping-pong balls -- radius $a =1.9$~cm and mass $m =2.7$~g -- that have been spray-painted silver to aid in visualization. Other systems have been analyzed but, except where noted, the diameter ratio of the large ball to the average diameter of the background is 4.4.

The full system -- background beads and one or two large balls -- is fluidized by an upflow of air at 300~cm/s, spatially and temporally homogeneous within $\pm10$~cm/s and 0.5~s, as verified by a hot-wire anemometer. The Reynolds number is $10^{4}$ such that the motion of the balls is driven stochastically by turbulence. The airflow is
below the terminal free fall speed of the balls, which ranges from 700-800~cm/s according to $mg = 0.43 (\rho_{air}/2)v^2 2 \pi r^2$,
so that the balls maintain contact with the sieve and move by rolling without slipping. As such, we define an effective mass $m_{\mathrm{eff}} = m + I/r^2$, where $I$ is the moment of inertia.

The apparatus, fluidization method, and lighting setup are
identical to those of Ref.~\cite{Ojha2005}. The apparatus is a rectangular windbox, $1.5\times1.5\times4$~ft.$^{3}$, positioned upright, with two nearly cubical chambers separated by a perforated metal sheet. A blower attached to the windbox base by a flexible cloth sleeve provides
vertical airflow perpendicular to a circular brass testing sieve with
mesh size 150~$\mu$m and radius $15.3$~cm that rests horizontally on
top. To prevent background spheres from getting caught in a small groove at the inner edge of the sieve, we place a 0.953~cm-diameter norprene tube around its inside edge. Thus, the system has an actual radius, $R = 14.3$~cm.

The particles are illuminated from above by six 100~W incandescent bulbs arranged in a 1-foot diameter ring positioned 3~feet above the sieve.
A digital CCD camera placed at the center of this ring captures the raw video data, typically for 4 hours at a time at 120 frames per second.
We threshold these long videos to binary as they save to
buffer so that only the highly-reflective large ball is seen. Post-processing of the video data is accomplished using LabVIEW, using custom particle-tracking programs. From the center-of-mass position obtained from the video data, we determine
velocity and acceleration by fitting the position data to a third-order polynomial over a window of $\pm 4$ frames. The window is Gaussian-weighted,
vanishing at the window edge, to ensure the continuity of the derivatives.   Scatter around the fits gives us an estimate of the error in the position data
of $\pm18~\mu$m.

We use the above process to track and characterize the background beads. The density of the background beads was chosen so that they were uniformly distributed across the system. We note that the packing is loose enough that the background bead dynamics are described by only a single ballistic-to-diffusive timescale and that there are no caging effects. From the short time mean square displacement, we can obtain the rms speed of the background beads as $v_{rms} = 1.18$~cm/s. This gives a value of $kT = m_{eff}\langle v^{2} \rangle/2 = 0.16~{\rm erg}$. For these conditions, the  the equation of state for air-fluidized beads~\cite{Daniels2010} gives the pressure, $P = 0.25~{\rm kg/s}^{2}$.  And the relaxation time is $\tau = 1.7$~s for the background beads to achieve an rms displacement comparable to their size.


\section{Single Ball Behavior}\label{oneball}
In this section we consider the behavior of a single large ball in a background of smaller beads. A single ping-pong fluidized in isolation behaves like a Brownian particle, obeying a thermal analogy~\cite{Ojha2005}. When fluidized in the presence of a background of smaller beads, the behavior is strikingly different. As soon as the large ball is placed in the background and fluidized, it begins to propel itself ballistically around the system. Typically, it will propel in a straight line until it reaches the boundary or the background balls jam in front of the large ball, at which point it is forced to change direction. When this occurs at the boundary, the particle begins to propel along the boundary edge resulting in circular motion at some stable orbit position. We emphasize the observed circular behavior in Figure~\ref{PhotoApparatus}(b). Here, we show a one-minute time-trace of the large ball center-of-mass position. Whenever the ball is moving clockwise, we color the trace red; counterclockwise motion is blue. Several circular orbits of both type can be seen in the figure. 

The mechanism by which the large ball propels is analogous to a previous experiment~\cite{Abate2005} where the size ratio of two balls was shown to be
a parameter by which one can progressively break the thermal behavior of the balls. In the case of two very differently-sized balls, when the two come into close contact with one another there is a short-range attraction which causes the two balls to stick together. They then propel, as a unit, along their line of centers in the direction of the smaller ball. We believe the same mechanism is at work in the current case of a large ball fluidized in a background of smaller balls.

In this experiment, the large ball receives a kick from the airflow or collides with a background ball and begins to move in some direction. The
background balls in front of the large ball become compressed whereas the region behind the large ball becomes dilute. This is because the large ball moves much faster than the small balls and the background is unable to rearrange itself quickly enough to fill in the void behind the large ball as it propels through the system. The dilute wake behind the ball is easily visible in Fig.~\ref{PhotoApparatus}(a).  The compressed region in front of the large ball behaves just like the smaller ball in the earlier two-ball experiment. The large ball moves in the direction of this compressed region and gains speed. 
This also suggests that the local fluctuations in the background density will be dependent on the speed of the large ball.

To quantify these observations, we begin by analyzing the effect of the large ball on the distribution of background balls; we then calculate time-independent probability distributions and the dynamics of the large ball.

\subsection{Background Beads}\label{background}
\begin{figure}[h!]
\begin{centering}
\includegraphics[width=0.9\textwidth]{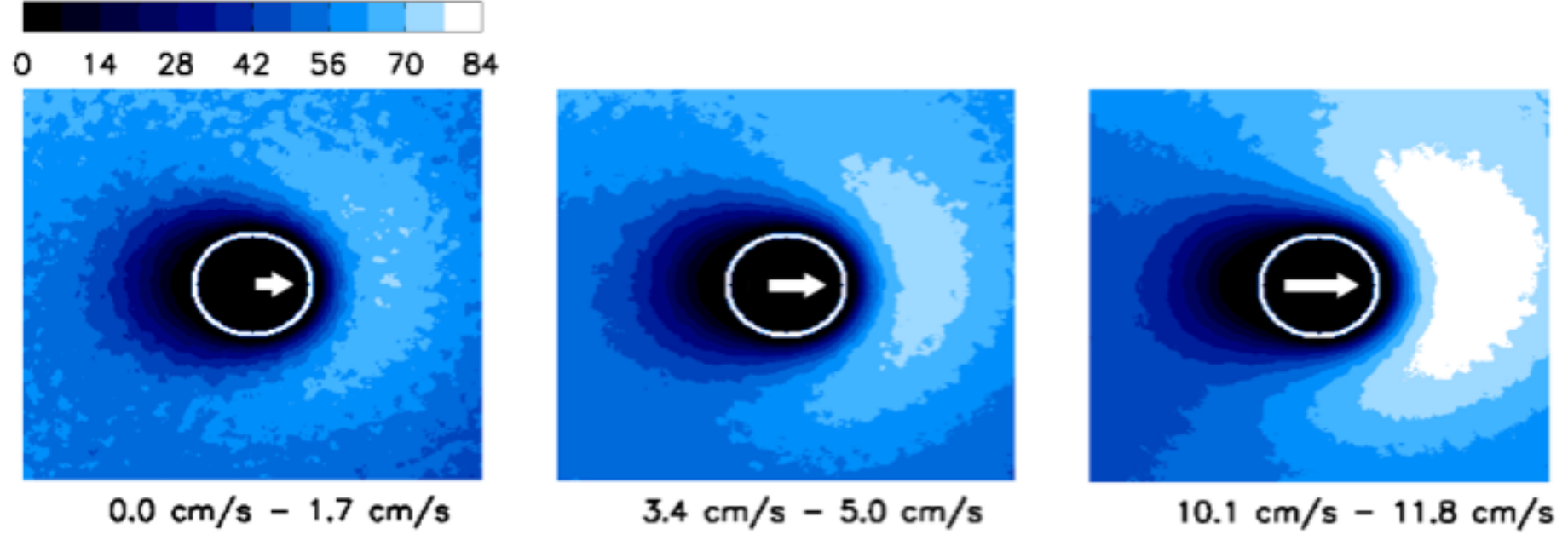}
\par\end{centering}
\caption{\label{BackgroundDensity} Time-averaged density of the background for three different ranges of the large ball speed, as specified below each image. The large ball (outlined in white), radius 1.9~cm, is always moving horizontally to the right. The density scale linearly interpolates between 0$\%$ area fraction (black) and a jammed region at 84$\%$ (white).}
\end{figure}

Rather than track the individual background beads, we quantify the local density fluctuations graphically. We first track the large ball position and then obtain its velocity in each frame. Then, each frame in the video was rotated and had its origin shifted such that the large ball was centered and moving to the right along the horizontal axis. Since we have suggested that the local density is dependent on the large ball speed, we bin these processed images according to the speed of the large ball and then average over all images within each speed bin. The results, for three different ranges of large ball speeds, are shown in Fig.~\ref{BackgroundDensity}. The images have been color-coded by linearly interpolating the grayscale values between 0$\%$ area fraction, shown as black, and jammed particles at 84$\%$, shown as white.

The effect is very dramatic. In the left image, the large ball -- outlined by a white circle -- is moving very slowly and the compressed and dilute
regions are relatively small. As we increase speed from left to right, we observe that both the compressed region in front of the ball and the dilute
wake behind the ball become larger. The slight asymmetry in the image is due to the ball circulating in the counterclockwise direction more often than clockwise for the video analyzed.

\begin{figure}[h!]
\begin{centering}
\includegraphics[width=0.7\textwidth]{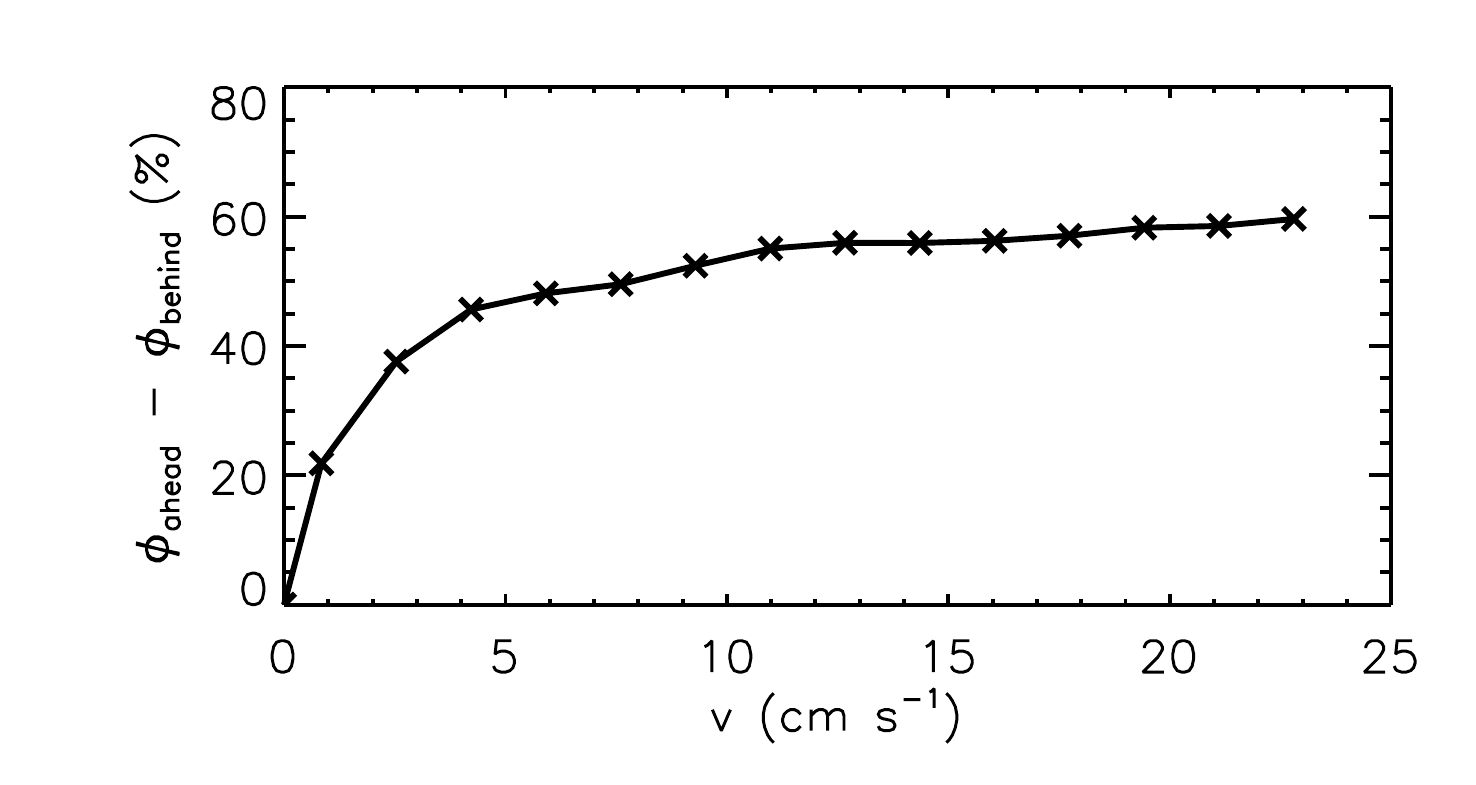}
\par\end{centering}
\caption{\label{SingleBallLambdaWithSpeed} The average difference in packing fraction of the background beads on opposites sides of a large fluidized ball as a function of its speed.  The beads are compressed in front and rarified in back, as seen in Fig.~\ref{BackgroundDensity}.}
\end{figure}

To quantify the difference in density between the compressed and rarefied regions, we obtain the average packing fraction within hemispherical annular areas both in front of, $\phi_{ahead}$, and behind, $\phi_{behind}$, the large ball.  We then plot the difference between these packing fractions as a function of the large ball speed, as shown in Fig.~\ref{SingleBallLambdaWithSpeed}.  The results show that the extent of the compression and rarefaction, and thus interaction of the large ball with the background, depends on the speed $v$ of the large ball.  The effect vanishes in the limit of zero speed, and saturates at $\phi_{\rm ahead}-\phi_{\rm behind}\approx0.6$ for $v>10$~cm/s.

\subsection{Statics}\label{statics}
Because our system reaches a steady state very quickly, the simplest way in which to characterize it is by compiling statistics of time-independent
quantities. Intuitively, we thought that the presence of the turbulent background would serve to overwhelm repulsive interactions of the large ball with the bounding walls. In other words, the large ball would move in a random walk, sampling all of the system space equally. In this case, the radial probability distribution $P(r)/r$ -- where we divide by $r$ to account for the fact that there are more points in phase space further from the origin -- would be constant from the origin $r = 0$ to the effective reduced radius of the system $R' = R-a =$ 12.4~cm, at which point the distribution would discontinuously vanish.

\begin{figure}[h!]
\begin{centering}
\includegraphics[width=0.9\textwidth]{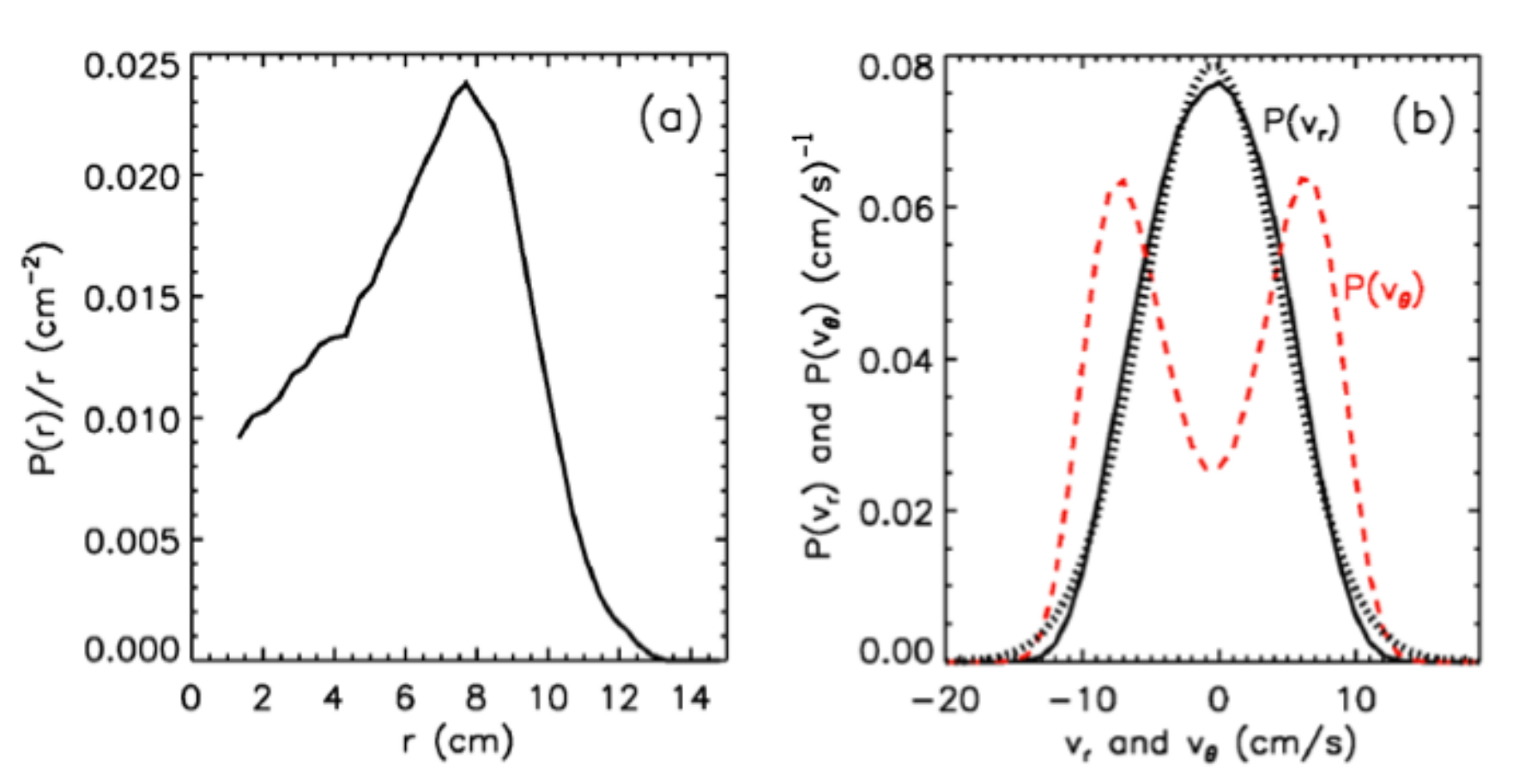}
\par\end{centering}
\caption{\label{vrandvtheta} (a) Radial probability distribution divided by radius, (b) radial velocity distribution (solid) fit by a Gaussian (dotted) and polar velocity (red dashed) distribution for a single large sphere fluidized in a background of smaller beads.}
\end{figure}

The observed $P(r)/r$ for our system is shown in Fig.~\ref{vrandvtheta}(a). In no region is the distribution constant, suggesting that the background may actually enhance the interaction with the boundary. There is a large peak at an intermediate radius as the large ball begins to detect the wall, showing that the large ball is repelled from the wall. This peak radius roughly corresponds to the radial position at which the particle prefers to orbit circularly.

Similarly, for a thermal particle obeying equipartition of energy, we would expect the velocity distribution $P(v)$ to be Gaussian, symmetric about zero, and that each velocity component would be equivalent. The compiled $P(v)$ for our system is shown in Fig.~\ref{vrandvtheta}(b). Here, we have decomposed the velocity into radial and polar components to emphasize the circular motion observed for this system.  The radial velocity distribution is roughly Gaussian, although having shorter tails. However, the polar velocity exhibits two peaks, consistent with circular behavior. The peaks are symmetric about the origin, showing that the particle does not preferentially orbit in any particular direction.

\subsection{Dynamics}\label{dynamics}
\begin{figure}[h!]
\begin{centering}
\includegraphics[width=0.7\textwidth]{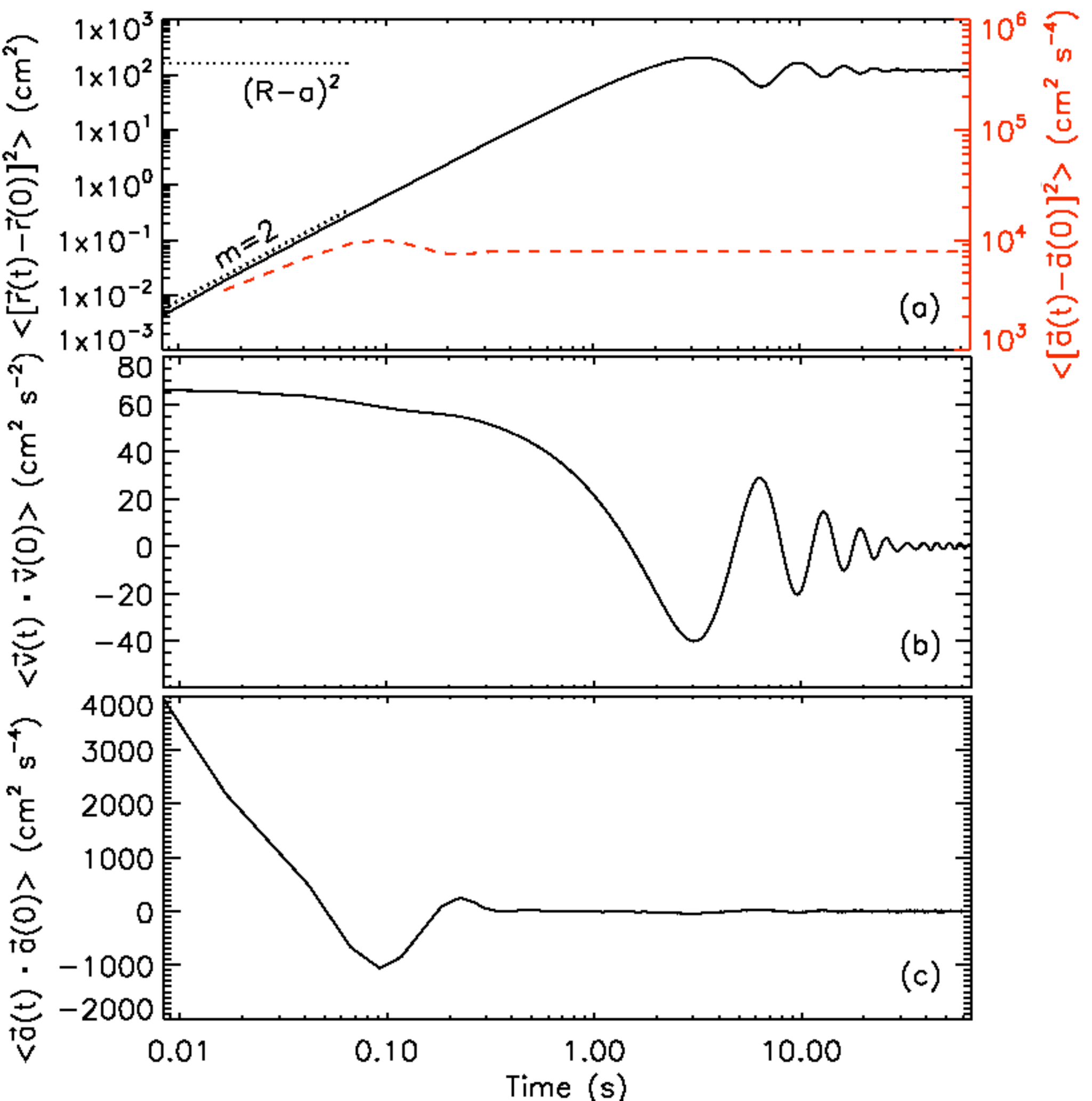}
\par\end{centering}
\caption{\label{SingleBallAutoCorrelations} (a) Mean square displacement (solid) and mean square acceleration fluctuations (red dashed), (b) velocity autocorrelation, and (c) acceleration autocorrelation for a single large sphere fluidized in a background of smaller beads.  The effective radius of the system is the difference $R'=R-a$ between the system and the large ball radii.}
\end{figure}

We further characterize the large ball behavior by quantifying its dynamics. As discussed above, the large ball propels itself ballistically around the system. This is directly observable in the mean-square displacement (MSD) for the large ball, shown in Fig.~\ref{SingleBallAutoCorrelations}(a). Here, at short times, we see ballistic motion ($\propto v_{rms}^{2}\tau^{2}$) characterized by a root-mean-square speed of $v_{rms}\approx4$~cm/s. By comparison, recall that tracking of individual background beads yield a root-mean-square speed of 1.18~cm/s that is about 3-4 times smaller. The system is too small for us to see any indication of a crossover to diffusive behavior. The MSD saturates within about 2~s at the optimum orbital radius, followed by oscillations about this value.

To get a sense of the characteristic timescales in our system, we consider the velocity and acceleration autocorrelation functions as shown in
Figs.~\ref{SingleBallAutoCorrelations}(b) and (c). The velocity autocorrelation has a long plateau that rolls off after approximately 1~s and then oscillates as the function decays to zero. This timescale corresponds roughly to the relaxation time of the large ball $a_{rms}/v_{rms}\approx0.95$~s. The strong oscillations are indicative of the circular motion exhibited by the ball. The acceleration autocorrelation decorrelates much more quickly than the velocity does, having one small oscillation at 0.08~s, before decaying to zero at approximately 0.3~s. This separation of timescales is further emphasized by including an overlay of the mean square acceleration change on the mean square displacement, Fig.~\ref{SingleBallAutoCorrelations}(a).


\subsection{Equation of Motion}\label{eqom}
In this section, we analyze the large ball motion in terms of a model of the forces acting on a single large ball fluidized in the presence of a background of smaller beads. A simple conceivable equation of motion for the large ball is
\begin{equation}
 m_{\mathrm{eff}} \vec{a} = C(r)\hat{r} + D(v)\hat{v} + \vec{\zeta}(t), \label{eqnmotion}
\end{equation}
where the forces $D(v)$ and $C(r)$ are functions of the large ball's speed and radial position, and $\vec{\zeta}(t)$ is a stochastic force satisfying $\langle\vec{\zeta}(t)\rangle = 0$. The position dependent force $C(r)\hat{r}$ must be circularly symmetric since the system has circular symmetry. The velocity dependent force must include a drag term for the interaction of the large ball with the mesh substrate, which we will assume to be linear since the velocity decays more slowly than the acceleration. As we saw in Section~\ref{background}, the interaction of the large ball with the background is also speed-dependent. Thus, we write $D(v)\hat{v} = -\gamma \vec{v} + \beta(v)\hat{v}$ where $-\gamma \vec{v}$ is the drag term and $\beta(v)\hat{v}$ is the force caused by the background which, for simplicity, we assume to be in the direction of the large ball velocity $\hat{v}$.

We will now examine each term of the equation using a dynamical approach. We can isolate the central force by taking the cross product of (\ref{eqnmotion}) with $\hat{v}$. Rearranging:
\begin{equation}  C(r) - \left[ \frac{\vec{\zeta} \times \hat{v}}{   \hat{r} \times \hat{v}} \right] =   \left( \frac{\vec{a} \times \hat{v}}{   \hat{r} \times \hat{v}}  \right) m_{\mathrm{eff}} \label{RadialForceEqn} \end{equation}
Similarly, we can isolate the speed-dependent force by taking the cross product of (\ref{eqnmotion}) with $\hat{r}$, giving:
\begin{equation}  D(v) - \left[ \frac{\vec{\zeta} \times \hat{r}}{   \hat{v} \times \hat{r}} \right] =   \left( \frac{\vec{a} \times \hat{r}}{   \hat{v} \times \hat{r}}  \right) m_{\mathrm{eff}} \label{SpeedForceEqn} \end{equation}

\begin{figure}[h!]
\begin{centering}
\centering
\includegraphics[width=1.0\textwidth]{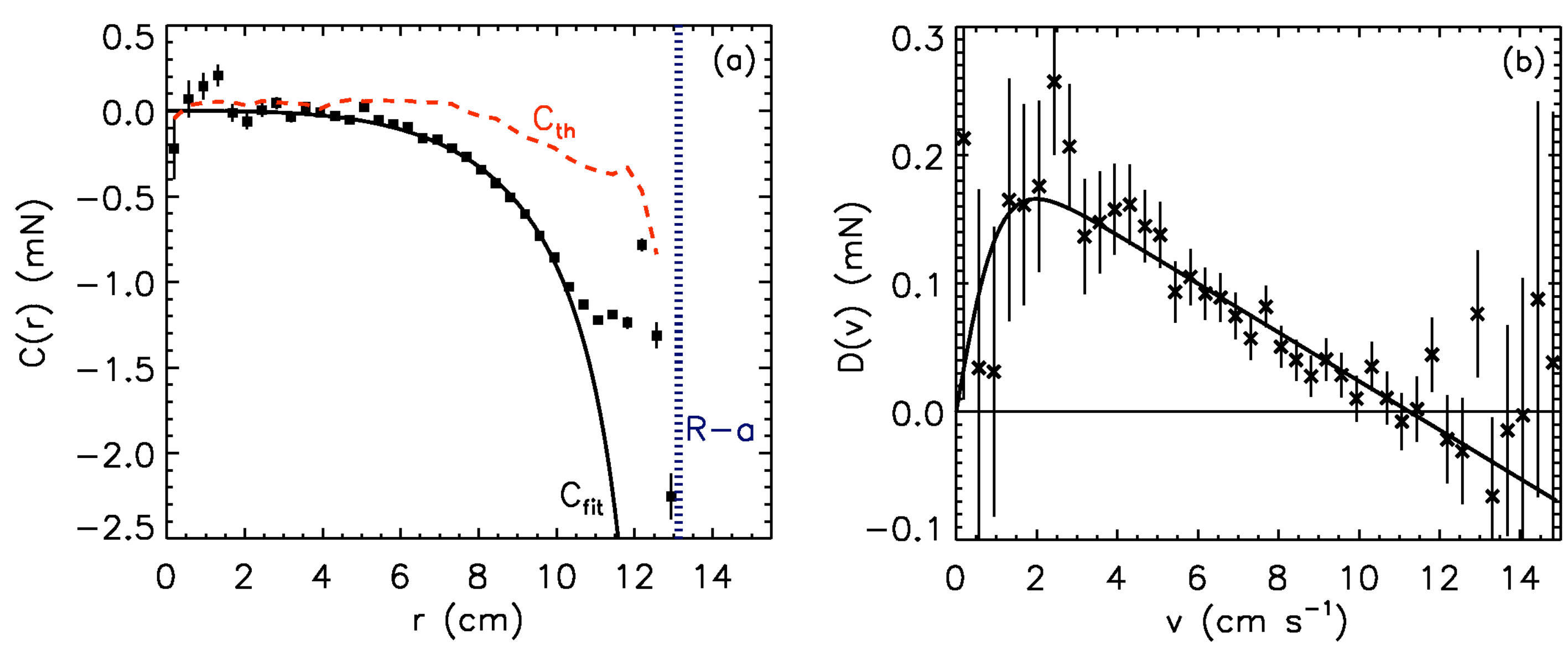}
\par\end{centering}
\caption{Forces acting on a large ball in a background of smaller beads: (a) central force $C(r)$ as given by (\ref{RadialForceEqn}) with a fit to $C(r)\propto r^{3}/(r-R')^{0.65}$ (solid) and the radial force obtained via thermal approximation according to $C_{th}(r)=-\frac{d}{dr}kT_{eff} \log(-P(r)/r)$ (red dashed); (b) speed-dependent force $D(v)$ as given by (\ref{SpeedForceEqn}) with an empirical fit to $D(v) = D_o \{ 2/[1+\exp{(-v/v_{\mathrm{switch}})}]-1\} - \gamma v$ (solid). The effective system boundary is marked at $r=R-a$ (vertical dotted blue line).
} \label{CentralPotentialWithOverlay}
\end{figure}

If we assume the stochastic force is independent of position and velocity, the time average of the square brackets in both (\ref{RadialForceEqn}) and (\ref{SpeedForceEqn}) must be zero. Thus, we can readily obtain $C(r)$, as shown in Fig.~\ref{CentralPotentialWithOverlay}(a). We see that there is essentially zero force at small radii. At increasing $r$, as the ball approaches the bounding walls, the force becomes repulsive. The solid line in the figure is a fit to $C(r)\propto r^{3}/(r-R')^{0.65}$. This is in contrast to the linear dependence of this term observed for a large sphere fluidized in an empty sieve~\cite{Ojha2004, Ojha2005}. As a further comparison, the dashed curve is the radial force obtained if we had approximated the system as thermal and calculated the radial force using the data for $P(r)$, according to $C(r)=-\frac{d}{dr}kT_{eff} \log(-P(r)/r)$.  The discrepancy between the two methods for deducing $C(r)$ shows that the system is not behaving thermally.

We can also readily obtain $D(v)$, as shown in Fig.~\ref{CentralPotentialWithOverlay}(b). For speeds between 0~cm/s and roughly 10~cm/s, the speed-dependent force is positive, causing the large ball to speed up. For speeds larger than about 10~cm/s, the force is negative thus slowing the particle.  This is consistent with our earlier observations of the compression/rarefaction that accompanies the large ball motion.  Once the ball
receives a kick and begins to move, it compresses a region of the background in front of it. This creates a feedback mechanism that causes the
particle to continue gaining speed. However, if the large ball moves too quickly, the compressed region becomes jammed and is able to slow the
large ball down.  Furthermore the polar velocity distributions in Fig.~\ref{vrandvtheta}(b) suggest that there is a stable polar speed of approximately 10~cm/s at which the particle orbits the system in circular motion. This is consistent with the speed at which $D(v)$ crosses the horizontal axis -- the stable speed at which there is no speed-dependent force. The solid curve shows a fit to the empirical form $D(v) =D_o \{ 2/[1+\exp{(-v/v_{switch})}] -1 \} - \gamma v$, where $v_{switch} =0.5$~cm/s, $\gamma = 1.9 \times 10^{-5}$~Ns/cm, and $D_o=0.2$~mN.

\begin{figure}[h!]
\begin{centering}
\centering
\includegraphics[width=1.0\textwidth]{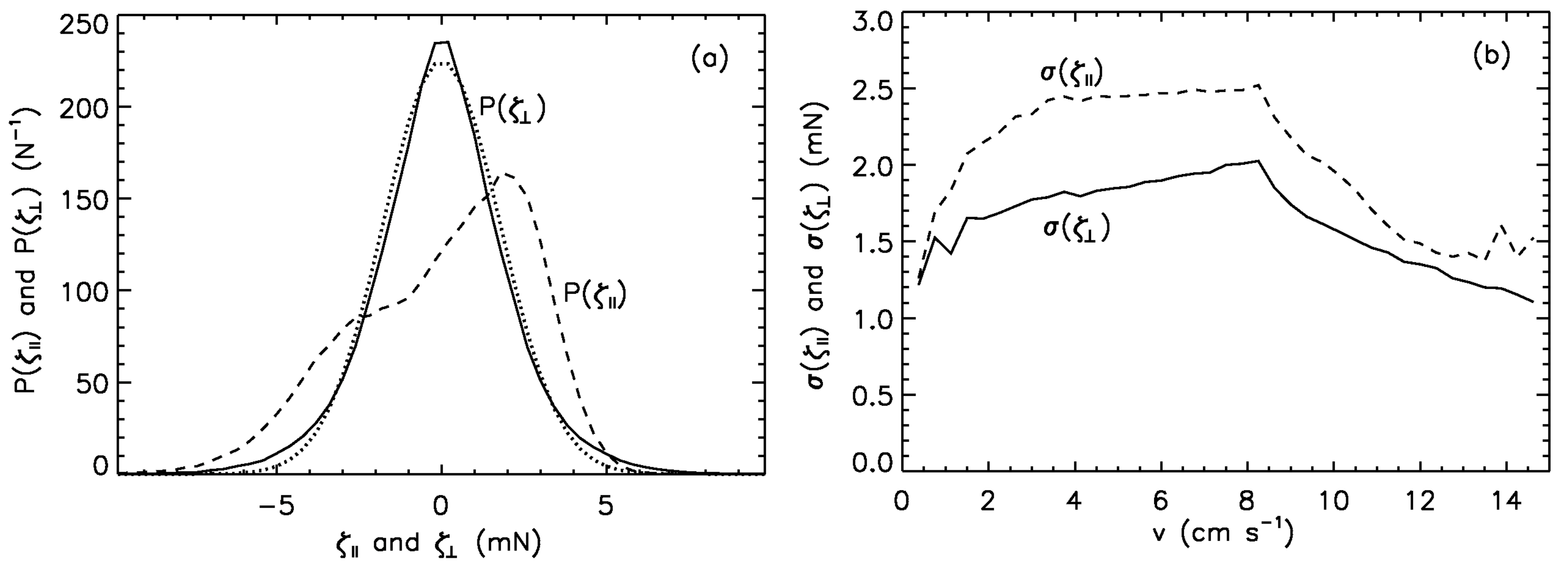}
\par\end{centering}
\caption{(a) Probability distribution of the stochastic force parallel to (dashed) and perpendicular to (solid) the large ball velocity, for $v$ = 4.5~cm/s.  The dotted curve is a Gaussian fit to $P(\zeta_\perp)$.  (b) Standard deviation of the stochastic force distributions as a function of large ball speed.}\label{dirnstochastic}
\end{figure}

Lastly, we examine the stochastic force $\vec{\xi}$. Our analysis thus far has assumed that the stochastic force is independent of the large ball position and velocity. Since the behavior of the background parallel to the large ball motion is quite different than perpendicular, we will analyze the stochastic force with respect to the direction of the large ball's motion. To do so, we first subtract the parameterized forces from (\ref{eqnmotion}) and take the dot and cross product with $\hat{v}$ to isolate the stochastic force parallel $\zeta_{\parallel}$ and perpendicular $\zeta_{\perp}$ to the large ball velocity:
\begin{equation} \zeta_{\parallel}(t) \equiv \vec{\zeta}\cdot \hat{v} = [m_{\mathrm{eff}} \vec{a} - C(r)\hat{r}] \cdot \hat{v} - D(v) \end{equation}
\begin{equation} \zeta_{\perp}(t) \equiv \vec{\zeta} \times \hat{v} = [m_{\mathrm{eff}} \vec{a} - C(r)\hat{r}] \times \hat{v}. \end{equation}
Since the speed directly enters these equations, we show the stochastic force probability functions in Fig.~\ref{dirnstochastic}(a) at the particular speed $v=4.5$~cm/s. $P(\zeta_{\perp})$, shown as the solid curve, is nearly Gaussian. By contrast, the
distribution of the parallel stochastic force is significantly skewed, having much larger probability to provide a kick in the direction of motion.
This shows that our assumption that the stochastic force is independent of the speed was incorrect. The shape of these distributions does not
significantly change depending on the magnitude of the large ball speed. This is verified in Fig.~\ref{dirnstochastic}(b) where we plot the standard
deviations $\sigma$ of the probability distributions $P(\zeta _{\parallel}(t))$ and $P(\zeta _{\perp}(t))$ as a function of the large ball speed. Both distributions remain roughly the same width until 8~cm/s, at which point the distributions narrow. Interestingly, this directional dependence of the stochastic force has been seen before in the behavior of fluidized rods~\cite{Daniels2009}. In that experiment, it was observed that the rods self-propel much like the large ball is observed to do in this experiment.

\section{Two Balls Behavior}\label{twoball}

Having documented the behaviour of a single large ball in a background of smaller balls, we add a second large ball to the system. When fluidized in an empty sieve, the force between two balls is characterized by hardcore repulsion and a persistent repulsive force over all separation distances~\cite{Ojha2005}. With the inclusion of background beads, we might expect a short-range depletion interaction that would attract the two balls together as well as longer-range forces mediated by the background. When we place the two large balls in the system, they both propel ballistically and orbit about the system. Often, the two balls will become trapped in one another's wake and travel as a pair. However, they are rarely observed to come into contact; when cooperatively traveling, the two balls are always some small distance apart.

\begin{figure}[h!]
\begin{centering}
\centering
\includegraphics[width=0.95\textwidth]{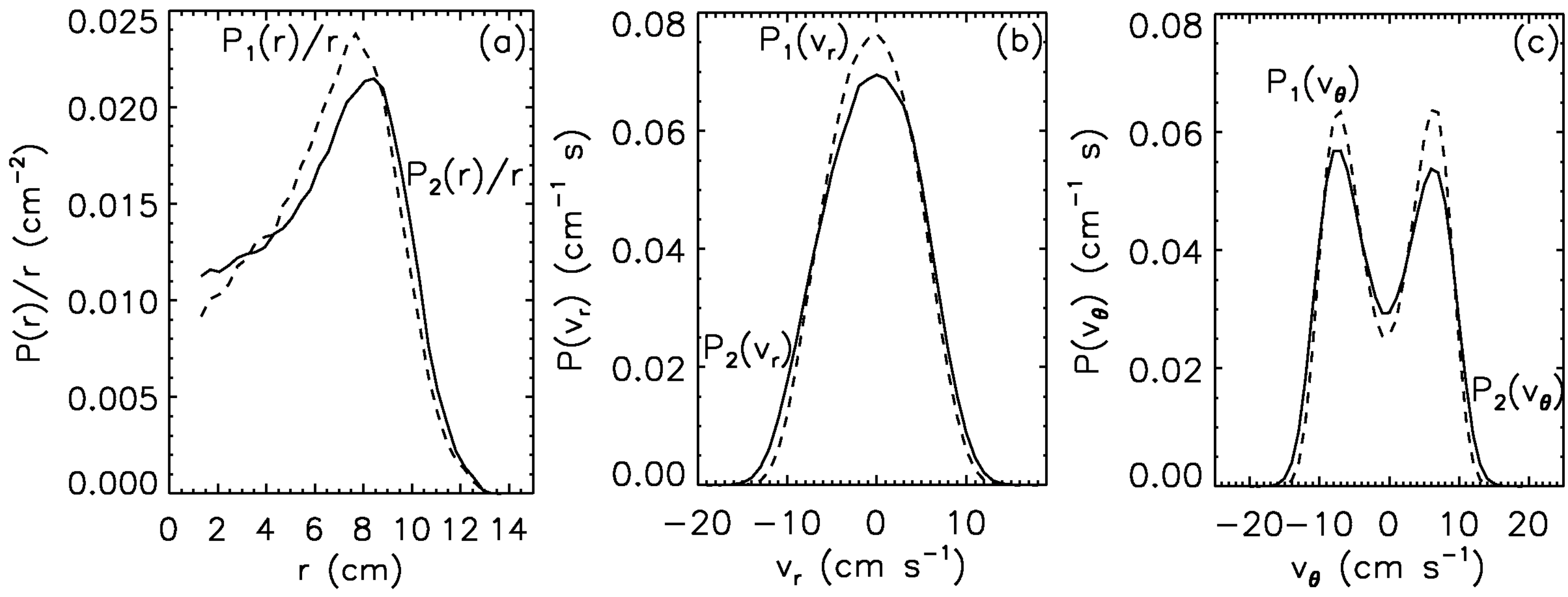}
\par\end{centering}
\caption{ Probability distributions for two large balls (solid line, subscript ``2''), compared with that for one lone large ball (dashed line, subscript ``1''), each in a background of smaller balls. (a) Radial distribution $P(r)/r$, (b) radial velocity distribution $P(v_r)$, and (c) theta velocity distribution $P(v_\theta)$. The similarity of the distributions with one and two large balls suggests an additional large ball does not drastically change the first ball's behaviour.}\label{ProbDistributionsTwoBalls}
\end{figure}

We begin by characterizing the behavior of the individual balls in the same way that we did for a single fluidized ball. The radial, radial velocity, and polar velocity probability distributions are shown in Figs.~\ref{ProbDistributionsTwoBalls}a-c. The distributions for the single large ball are shown as dashed curves to highlight differences. The distributions are qualitatively unchanged, suggesting the behaviour of one large ball in a background of smaller balls is not greatly affected by another large ball.

\begin{figure}[h!]
\begin{centering}
\centering
\includegraphics[width=0.8\textwidth]{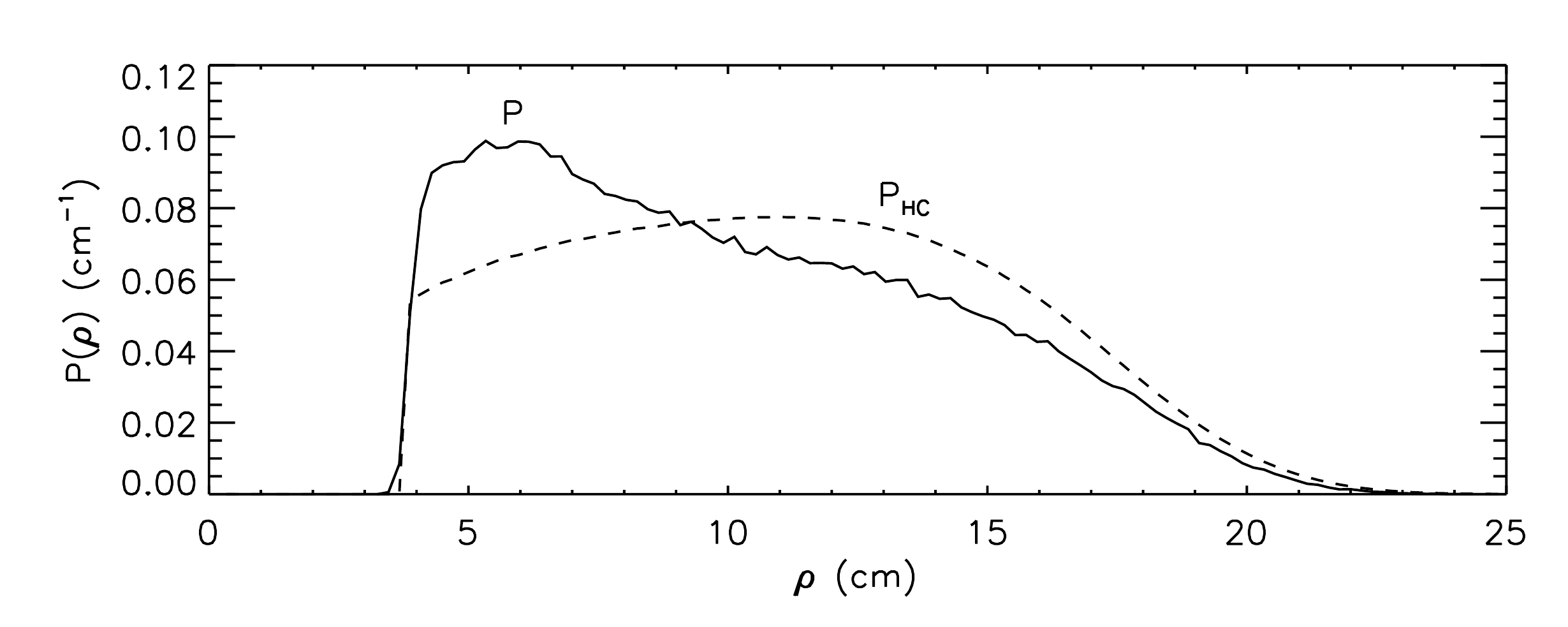}
\par\end{centering}
\caption{ Separation distribution $P(\rho)$ for the separation $\rho$ between two large balls in a background of smaller balls (solid) and simulation results $P_{HC}(\rho)$ (dashed) for balls interacting only via hard core repulsion.}\label{ProbDistributionsTwoBallsSep}
\end{figure}

Next we compile the probability distribution for separation distance,
$P(|\vec{r}_1-\vec{r}_2|) = P(\rho)$ vs $\rho$. This is shown as the solid line in Fig.~\ref{ProbDistributionsTwoBallsSep}. The distribution is zero for separations less than a particle diameter 3.8~cm, as expected for hard spheres. At small separations, the
probability reaches a maximum showing that the balls prefer strongly to stay near one another. This immediately suggests that the nature of the
interaction between the two balls has dramatically changed and become attractive.  However, we note that depletion
interactions act on a length scale of the size of the small balls $\sim$ 0.87~cm. The broadness of the peak indicates that the interaction
is over a length scale of approximately 5~cm and cannot be attributed to depletion interactions. In fact, the absence of a strong peak within one
small-ball diameter from the large ball diameter suggests that depletion interactions play no significant role here.  To stress the contribution due to the background, we generate the distribution $P_{\mathrm{HC}}(\rho)$ expected for hard core repulsion alone, shown as the dashed curve, using a Monte Carlo simulation in which pairs of positions were generated randomly for two fictitious balls by randomly choosing a phase and radial position weighted by $P_{1}(r)$ data for each. As such, any difference between our data and the simulation is due to interactions either mediated by the background beads or by the airflow, which is know to be repulsive \cite{Ojha2005}. The comparison shows an enhancement of the distribution at close separation, which suggests an effective attraction between the large balls mediated by the background beads.   We further emphasized that this attractive force is mediated by the background by anchoring one of the large balls to be stationary while the other large ball freely moves about the system.  The resulting separation probability distribution is identical to the one-ball radial probability distribution $P_{1}(r)$.  Therefore, the attractive force between two free balls is dependent on their motion through the background and presumably linked to fluctuations in the local density.

\section{Ball-Ball Interaction}\label{interact}
\begin{figure}[h!]
\begin{centering}
\centering
\includegraphics[width=1.0\textwidth]{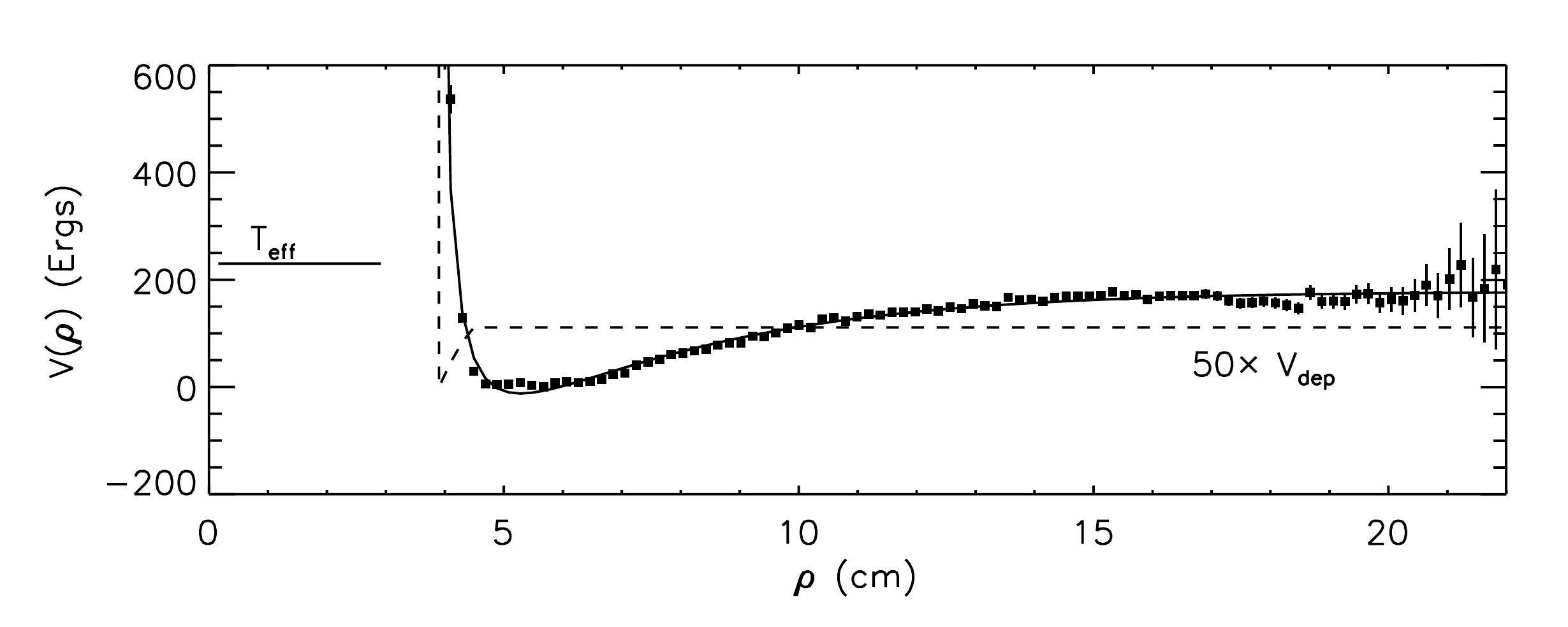}
\par\end{centering}
\caption{ Interaction potential for two large balls in a background of smaller beads, calculated from the measured separation distribution $P(\rho)$ and assuming thermal equilibrium. The solid line is a fit to $V_s^{\mathrm{fit}}(\rho) = V_\alpha[(\rho - 2a)/a]^\alpha + V_0\exp[-(\rho - 2a)/\lambda]$. The dashed line shows the depletion potential $V_{\mathrm{dep}}(\rho)$, multiplied by a factor of 50.}\label{TwoBallsThermalInterForce}
\end{figure}

In this section we seek to quantify the interaction force between two balls in a background of smaller beads. We assume the same equation of motion as for one large ball, but add interaction forces:
\begin{eqnarray}
m_{\mathrm{eff}} \vec{a}_1 &=&  C(r_1)\hat{r}_1 + D(v_1)\hat{v}_1 + \vec{\zeta}_1(t) + \vec{F}_{1\,2} \label{eqnmotionball1} \\
m_{\mathrm{eff}} \vec{a}_2 &=&  C(r_2)\hat{r}_2 + D(v_2)\hat{v}_2 + \vec{\zeta}_2(t) + \vec{F}_{2\,1}
\end{eqnarray}
Assuming $\vec{F}_{1\,2}$ depends only on $\rho$, we define an interaction potential $V_s(\rho)$ according to
\begin{equation}
 \vec{F}_{1\,2}(\rho) = [-\mathrm{d}V_s(\rho)/\mathrm{d}\rho] \hat{\rho}_{1\,2} ~~~~~~~~~~\mathrm{where}~~ \hat{\rho}_{1\,2} = (\vec{r}_1-\vec{r}_2)/|\vec{r}_1-\vec{r}_2|
\label{PotentialDefinition}
\end{equation}
First, we want to determine the interaction potential $V_{s}(\rho)$ from our data. We can infer this from the separation distribution $P(\rho)$ if we assume the system behaves sufficiently thermally such that
\begin{equation}
 P(\vec{r}_1,\vec{r}_2) \propto \exp{\left\{-[V_c(r_1) + V_c(r_2) + V_s(\rho)]/T_{\mathrm{eff}} \right\} } \label{ThermalProbability}
\end{equation}
where the central potential for one-large-ball is given by $ V_c(r)= - T_{\mathrm{eff}} \ln(P_1(r)/r)$. Rearranging (\ref{ThermalProbability}) and integrating over remaining variables, $V_{s}(\rho)$ can be obtained from the measured $P(\rho)$:
\begin{equation}
V_{s}(\rho) =  - T_{\mathrm{eff}} \ln[P(\rho)/g(\rho)] + \mathrm{const.} \label{ProbDensityEqn}
\end{equation}
where
\begin{equation}
 g(\rho) = \rho \int_0^{R-a} \mathrm{d} r_1 \int_{0}^{2 \pi} \mathrm{d} \varphi ~r_1~ \exp{\left[-\frac{(V_c(r_1)+V_c(r_2))}{ T_{\mathrm{eff}}} \right]}. \label{Gdetermination}
\end{equation}
The resulting interaction potential is shown as the points in Fig.~\ref{TwoBallsThermalInterForce}. As expected due to hard-core repulsion, the potential diverges at one ball diameter. The data has been fit to the empirical form $V_s^{\mathrm{fit}}(\rho) = V_\alpha[(\rho - 2a)/a]^\alpha + V_0\exp[-(\rho - 2a)/\lambda]$.

It is straightforward to show that the data is inconsistent with a depletion attraction. A depletion interaction between large balls is expected when they are sufficiently close that small balls cannot access the region between them. The background beads impart a constant pressure $p$ on all accessible surfaces. This pressure $p$ can be estimated by assuming that the background beads obey a reduced-volume ideal gas equation of state $p = N_{b} T_{\mathrm{b}}/A_{reduced}$ where $ T_{\mathrm{b}}=m_{\mathrm{b,eff}} \langle  v_{\mathrm{b}}^2 \rangle$ and the reduced area $A_{reduced}$ is the area available to the balls' centers, minus the minimum area they could occupy. 
By integrating over the exposed circumference of one of the large balls, we can derive the depletion force $F_{\mathrm{dep}}(\rho) \hat{\rho}_{1\,2}$ on ball 1 due to ball 2. This force will be non-zero and finite only for the region $2a < \rho < 2(a+b)$, where $a$ is the large ball radius and $b$ is the average background bead radius. 
In this region, the force is given as $F_{\mathrm{dep}}(\rho)= -2 a p \sqrt{1 - \left[\rho/(2a+2b) \right]^2 } $. Then, we are able to obtain the interaction potential by integration:
\begin{equation}
 V_{\mathrm{dep}}(\rho) = -2 a p  \left\{     \frac{\rho}{2} \sqrt{1 - \left[ \frac{\rho}{2(a+b)} \right]^2 } + (a+b) \sin^{-1}\left[ \frac{\rho}{2(a+b)} \right] \right\}.
\label{DepletionPotential}
\end{equation}
This expression has been plotted alongside our data as the dashed curve in Fig.~\ref{TwoBallsThermalInterForce}, multiplied by a factor of 50. Both the range and strength of a depletion attraction are much smaller than that obtained from the data.

\begin{figure}[h!]
\begin{centering}
\centering
\includegraphics[width=1.0\textwidth]{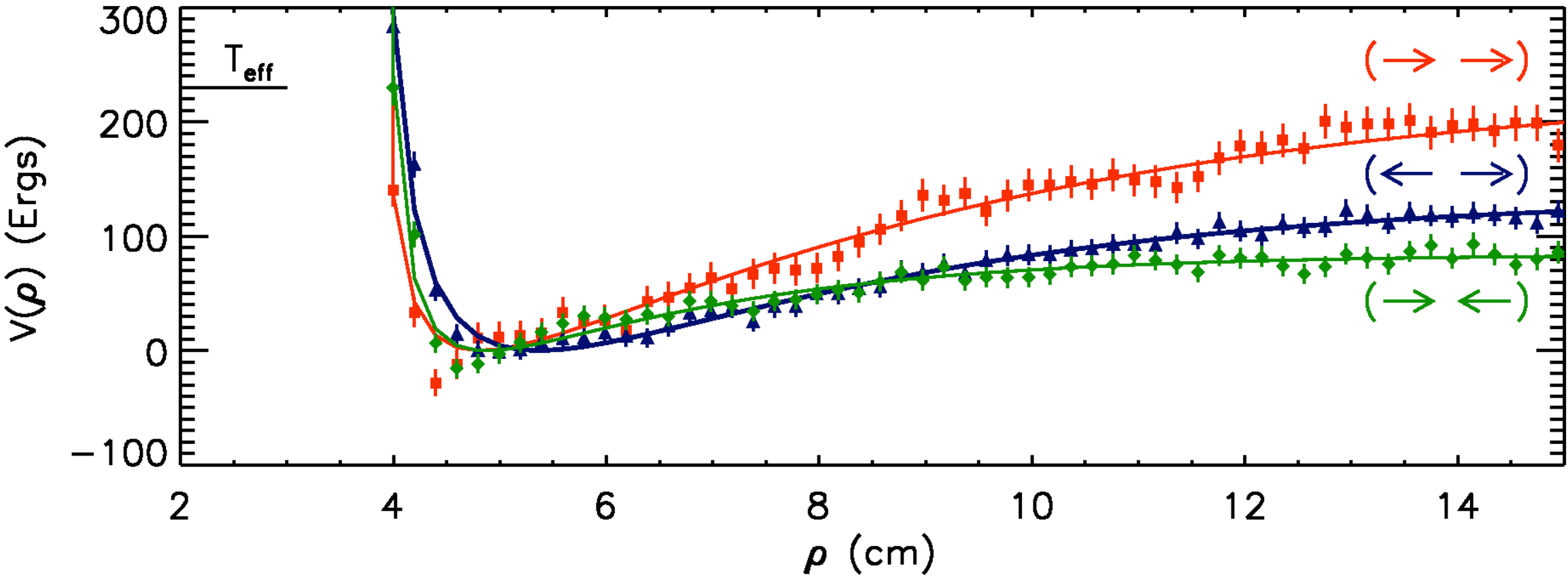}
\par\end{centering}
\caption{Interaction potentials for two large balls in a background of smaller beads for different relative motion of the large balls, calculated from the measured separation distribution $P(\rho)$ and assuming thermal equilibrium: moving in the same direction ($\rightarrow \rightarrow$, red squares); moving apart ($\leftarrow \rightarrow$, blue triangles); and moving closer together ($\rightarrow \leftarrow$, green diamonds). The solid curves are best fit lines to the empirical form $V_s^{\mathrm{fit}}(\rho) = V_\alpha[(\rho - 2a)/a]^\alpha + V_0\exp[-(\rho - 2a)/\lambda]$.}\label{TwoBallsThermalInterForceTogApart}
\end{figure}

Just as the force on a single ball is speed-dependent, it is likely that the ball-ball interaction is also dependent on how the balls are moving relative to one another. To examine this, we first separated the video data into three scenarios of large ball motion:
\begin{enumerate}
 \item $(\rightarrow \rightarrow)$: moving in same direction [i.e. $\vec{v}_1 \cdot \vec{v}_2>0$]
 \item $(\leftarrow \rightarrow)$: moving apart [i.e. $\vec{v}_1 \cdot \vec{v}_2<0$ and $\mathrm{d}\rho/\mathrm{d}t >0$]
 \item $(\rightarrow \leftarrow)$: moving closer together [i.e. $\vec{v}_1 \cdot \vec{v}_2<0$ and $\mathrm{d}\rho/\mathrm{d}t <0$]
\end{enumerate}
We then repeat the above analysis to determine the interaction potential for different types of relative motion. Our results are shown in Fig.~\ref{TwoBallsThermalInterForceTogApart}. The potentials are significantly different from each other, which is strong evidence that the interaction force depends on the velocity of the large balls as well as their separation. Each scenario displays short range repulsion with attraction at longer ranges. The strongest attraction occurs when the balls move in the same direction.

\section{Conclusions}
We have characterized the behavior of and forces acting on a large sphere fluidized in a bidisperse background of smaller beads. The large ball self-propels ballistically through the medium, strongly perturbing the local density of the background. The presence of the background not only modifies the central force felt by a fluidized ball moving in an empty sieve but also mediates a novel speed-dependent force. The speed-dependent force acts to keep the large ball propelling itself at a stable speed of approximately 10~cm/s. Furthermore, the stochastic force is directionally-dependent and preferentially kicks the large ball in its direction of motion.  When two large balls are fluidized simultaneously, there is a short-range repulsion mediated by the airflow, as in Ref.~\cite{Ojha2005} but with a range that is cut off.  And there is a long range attraction mediated by the background beads and dependent on the relative motion of the two balls.   Further, we demonstrated that the attraction is not consistent with the depletion interaction.

\section{Acknowledgments}
This work was supported by the NSF through grant DMR-0704147.

\section*{References}
\addcontentsline{toc}{part}{References}
\bibliographystyle{spphys}
\bibliography{reference}

\begin{thebibliography}{10}
\providecommand{\url}[1]{{#1}}
\providecommand{\urlprefix}{URL }
\expandafter\ifx\csname urlstyle\endcsname\relax
  \providecommand{\doi}[1]{DOI \discretionary{}{}{}#1}\else
  \providecommand{\doi}{DOI \discretionary{}{}{}\begingroup
  \urlstyle{rm}\Url}\fi

\bibitem{Reulle2004}
D.~Ruelle, Physics Today \textbf{57}(5), 48 (2004)

\bibitem{CMMP2010}
CMMP-2010, \emph{Condensed-Matter and Materials Physics: The Science of the
  World Around Us} (The National Academies Press, Washington, DC, 2007)

\bibitem{Nedderman}
R.M. Nedderman, \emph{Statics and Kinematics of Granular Materials} (Cambridge
  Univ. Press, NY, 1992)

\bibitem{BehringerRev}
H.M. Jaeger, S.R. Nagel, R.P. Behringer, Rev. Mod. Phys. \textbf{68}, 1259
  (1996)

\bibitem{Duran}
J.~Duran, \emph{Sands, powders, and grains} (Springer, NY, 2000)

\bibitem{Kubo1991}
R.~Kubo, M.~Toda, N.~Hashitsume, \emph{Statistical Physics II: Nonequilibrium
  Statistical Mechanics} (Springer, NY, 2001)

\bibitem{Clark1990}
B.~Pouligny, R.~Malzbender, P.~Ryan, N.A. Clark, Phys. Rev. B \textbf{42}, 988
  (1990)

\bibitem{Bideau1995}
I.~Ippolito, C.~Annic, J.~Lemaitre, L.~Oger, D.~Bideau, Phys. Rev. E
  \textbf{52}, 2072 (1995)

\bibitem{UrbachPRE99}
J.S. Olafsen, J.S. Urbach, Phys. Rev. E \textbf{60}, R2468 (1999)

\bibitem{PrentissAJP00}
J.~Prentis, Am. J. Phys. \textbf{68}, 1073 (2000)

\bibitem{Baxter2003}
G.W. Baxter, J.S. Olafsen, Nature \textbf{425}, 680 (2003)

\bibitem{Danna2003}
G.~D'Anna, P.~Mayor, A.~Barrat, V.~Loreto, F.~Nori, Nature \textbf{424}, 909
  (2003)

\bibitem{Ojha2004}
R.~Ojha, P.A. Lemieux, P.K. Dixon, A.J. Liu, D.~Durian, Nature \textbf{427},
  521 (2004)

\bibitem{SongMakse05}
C.M. Song, P.~Wang, H.A. Makse, PNAS \textbf{102}, 2299 (2005)

\bibitem{Abate2005}
A.R. Abate, D.J. Durian, Phys. Rev. E \textbf{72}, 031305 (2005)

\bibitem{Abate2008}
A.R. Abate, D.J. Durian, Phys. Rev. Lett. \textbf{101}, 245701 (2008)

\bibitem{MaksePRE08}
P.~Wang, C.~Song, C.~Briscoe, H.A. Makse, Phys. Rev. E \textbf{77}, 061309
  (2008)

\bibitem{UrbachPRE07}
P.~Melby, A.~Prevost, D.A. Egolf, J.S. Urbach, Phys. Rev. E \textbf{76}, 051307
  (2007)

\bibitem{Ojha2005}
R.P. Ojha, A.R. Abate, D.J. Durian, Phys. Rev. E \textbf{71}, 016313 (2005)

\bibitem{Daniels2010}
L.J. Daniels, T.K. Haxton, N.~Xu, A.J. Liu, D.J. Durian.
\newblock In preparation

\bibitem{Daniels2009}
L.J. Daniels, Y.~Park, T.C. Lubensky, D.J. Durian, Phys. Rev. E \textbf{79},
  041301 (2009)

\end{thebibliography}
\end{document}